\documentclass[10pt,letterpaper]{article}
\usepackage{opex3}
\usepackage{color}

\usepackage{caption}
\usepackage{verbatim}
\usepackage{subfig}
\usepackage{amsmath}
\usepackage[toc,page]{appendix}
\usepackage{epstopdf}
\usepackage{cite}
\usepackage{caption}
\usepackage{subfig}

\usepackage{breqn}

\newcommand{\Wmap}{W_{\rm{re}}(z)}
\newcommand{\mR}{m}

\begin{document}

\title{Energy density distribution of shaped waves inside scattering media mapped onto a complete set of diffusion modes}

\author{Oluwafemi S. Ojambati$^{1,*}$, Allard P. Mosk$^1$, Ivo M. Vellekoop$^2$, Ad Lagendijk$^1$, and Willem L. Vos$^1$}

\address{$^1$Complex Photonic Systems (COPS), MESA+ Institute for Nanotechnology, University of Twente, PO Box 217, 7500 AE Enschede, The Netherlands\\
$^2$ MIRA Institute for Biomedical Technology and Technical Medicine, University of Twente, PO Box 217, 7500 AE Enschede, Netherlands}

\email{$^*$o.s.ojambati@utwente.nl} 



\begin{abstract}
We show that the spatial distribution of the energy density of optimally shaped waves inside a scattering medium can be described by considering only a few of the lowest eigenfunctions of the diffusion equation. 
Taking into account only the fundamental eigenfunction, the total internal energy inside the sample is underestimated by only 2$\%$.
The spatial distribution of the shaped energy density is very similar to the fundamental eigenfunction, up to a cosine distance of about 0.01.
We obtained the energy density inside a quasi-1D disordered waveguide by numerical calculation of the joined scattering matrix. 
Computing the transmission-averaged energy density over all transmission channels yields the ensemble averaged energy density of shaped waves. 
From the averaged energy density obtained, we reconstruct its spatial distribution using the eigenfunctions of the diffusion equation. 
The results from our study have exciting applications in controlled biomedical imaging, efficient light harvesting in solar cells, enhanced energy conversion in solid-state lighting, and low threshold random lasers.
\end{abstract}



\section{Introduction}
There are numerous scattering media that are either of natural origin - such as biological tissue, fog, cloud, and teeth - or man-made - such as paint, diffuser glass, and phosphor inclusion in white LEDs.
These materials multiply scatter waves in all directions, due to the random spatial variations of the refractive indices.
Random interference occurs between the multiply scattered waves that is well-known as speckles~\cite{Goodman2000}. 
Upon averaging over realizations of scatterers, the speckles average out and the resulting average energy density is well described by diffusion theory~\cite{ishimaru1978Book,vanRossum1999RevModPhys,akkermans2007book}. 

Recently, novel wave-shaping methods such as feedback-based wavefront shaping \cite{freund1990PhysicaA,vellekoop2007OptLett,vellekoop2008PRL, popoff2014PRL,Davy2012PRB,mosk2012NatPhoton,vellekoop2015OptExp}, time reversal\cite{Derode1995PRL,lerosey2004PRL,lerosey2007Science}, phase conjugation\cite{leith1966JOSA,Dowling1992JourAcouAm,yaqoob2008NatPhoton}, and transmission matrix-based control \cite{kim2012NatPhoton,Popoff2010PRL,popoff2011NJP} have demonstrated the control of transmission and reflection through scattering media. 
These wave-shaping methods remarkably enhance the intensity in a single speckle as well as the total reflected and transmitted intensity.
The wave-shaping methods have led the way for exciting applications such as non-invasive biomedical imaging~\cite{Wang2012NatComm,Si2012NatPhoton,Bertolotti2012Nature}, advanced optics~\cite{Cizmar2010NatPhot,ParkOptExp2012,GuanOptLett2012,Park2012OptLett,SmallOptLett2012,Huisman2015OptExp}, and cryptography and secure communication\cite{Horstmeyer2013SciReport, Goorden2014Optica}. 

Although these methods are useful tools in controlling the intensity at the interfaces of a scattering medium, the energy density distribution of the shaped waves inside the scattering medium is still unknown.
In the case of light, the knowledge of the energy density distribution is important for applications such as enhanced energy conversion in white LEDs~\cite{krames2007JDispTechnol,phillips2007LaserPhotonRev,Leung2014OptExp,ogi2013ECIJSolidStateSciTechnol,Meretska16arXiv}, efficient light harvesting in solar cells~\cite{levitt1977ApplOpt,polman2012Nature,Si2014APL}, and controlled illumination in biomedical imaging~\cite{yizhar2011Neuron}. 
To date, only numerical calculations of scalar waves~\cite{choi2011PRB, davy2015NatComm, liew2015OptExp} and a single-realization elastic wave experiment~\cite{aubry2014PRL} have addressed the energy density distribution of shaped waves inside two-dimensional (2D) scattering media. 
However, none of these studies provide an analytical model for the energy density distribution.
For high-transmission channels, Davy \textit{et al}~\cite{davy2015NatComm} described the energy density using a parabolic function.
In Ref.~\cite{OjambatiArxiv}, our team reported the first measurement of the total energy density of light inside a three-dimensional (3D) scattering medium and reported an enhanced total energy density for shaped waves. 
The results were interpreted with a heuristic model that the total energy density of shaped waves can be described with only the fundamental eigenfunction of the diffusion equation, which agreed well with experimental observation.
However, the exact distribution of the spatial energy density for shaped waves still remains unknown.

In this paper, we study the energy density of shaped waves inside a scattering medium by using the eigenfunctions of the diffusion equation to reconstruct the energy density distribution. 
Here, we refer to shaped waves to mean perfect phase conjugation of the transmitted waves, which is equivalent to shaping to optimally focus light to a single speckle spot. 
The diffusion eigenfunctions form a complete set and therefore describe the ensemble average energy density within the domain of the sample boundary. 
By numerical calculation of concatenated scattering matrices~\cite{Ko1988PRB,Pendry1992PRL,Beenakker1997RMP}, we obtain the intensity transmission coefficients $\tau_n$ of $n$th transmission channel and its energy density $W_n(z,\tau)$ along the depth $z$ of a quasi-1D disordered waveguide.  
A quasi-1D disordered waveguide provides a tractable platform to study the essential physics.
By calculating the transmission-averaged energy density for all transmission channels, we obtained the energy density of shaped waves~\cite{vellekoop2008PRL,vellekoop2015OptExp}. 
We show that only the first seven diffusion eigenfunctions are sufficient to reconstruct the spatial energy density distribution of the shaped waves.
Taking into account only the fundamental eigenfunction of the diffusion equation, the total internal energy inside the sample is underestimated by only 2$\%$.
The spatial energy density is very similar to the fundamental eigenfunction, with a cosine distance of 0.01.
Furthermore, we are able to reconstruct the energy density distribution of both high- and low- transmission channels, using a few $M$ eigenfunctions of the diffusion equation, \textit{e.g.} $M = 16$ for channel with transmission $\tau = 0.1$.
The reconstruction of the distribution of the energy densities that we found here can be extended to three dimensional (3D) scattering samples as well as to other forms of classical and quantum waves. 

\section{Theory}

\subsection{Eigenfunctions of the diffusion equation}
The diffusion equation describes the ensemble averaged energy density $W(\textbf{r},t)$ of multiply scattered waves as a function of position $\textbf{r}$ and time $t$ inside a scattering medium~\cite{ishimaru1978Book, vanRossum1999RevModPhys},
\begin{equation}
\frac{\partial W(\textbf{r},t)}{\partial t} = D \nabla^2 W(\textbf{r},t) ,
\label{eq:dif3dt}
\end{equation}
where $D$ is the diffusion constant. Anticipating the fact that the directions $(x,y)$ are decoupled from the $z$-direction we decompose the energy density as
%
\begin{equation}
W(\textbf{r},t) =  W_\perp(x,y,t)\  W(z,t) \, .
\label{eq:W_xyz}
\end{equation}
Since we are considering a slab that is a quasi-1D scattering medium, we assume translational invariance in the perpendicular directions $(x,y)$. 
The sample boundaries are at $z = 0$ and $z = L$, where $L$ is the sample thickness.
This symmetry allows us to solve $W_\perp(x,y,t)$ by using its two-dimensional spatial Fourier transform $W_\perp(\textbf{q}_\perp,t)$. Solving the $(x,y)$-part of Eq.~(\ref{eq:dif3dt}) gives
\begin{equation}
W_\perp(\textbf{q}_\perp,t)
 = W(\textbf{q}_\perp,t=0)\ e^{-Dq_\perp ^2 t} \, .
\label{eq:W_xy}
\end{equation}

Next we solve the remaining 1D equation
\begin{equation}
\frac{\partial W(z,t)}{\partial t} = D \frac{\partial^2 W(z,t)}{\partial z^2}
\label{eq:diff1dt}
\end{equation}

We will turn this into a Sturm-Liouville eigenvalue problem by introducing the Ansatz
$W(z,t) = e^{-\gamma t} W(z)$
\begin{eqnarray}
 \frac{\partial^2 W(z)}{\partial z^2} + \Gamma \ W(z)&=&0 \, ,
\label{eq:Sturm-Liouville} \\
\Gamma &\equiv& \frac{\gamma}{D}\, .
\end{eqnarray}
As is well-known from linear algebra \cite{Morse1953Book} solutions of Sturm-Liouville equations are orthogonal and form a complete set. 
Applying boundary condition will lead to a discrete set of eigenvalues {$\Gamma_m$}. The general eigenfunction for eigenvalue $\Gamma_m$ can be cast in the form
\begin{equation}
W_m(z) =  A_\mR^{-1/2}\sin (\kappa_m z + \eta_\mR) \, .
\label{eq:W_modes}
\end{equation}
As outlined in detail in Ref. \cite{lagendijk1989PLA} the wavevector $\kappa_m$ and phase $\eta_m$ will be determined from solving Eq.~\ref{eq:Sturm-Liouville} with the appropriate boundary conditions. The paramater $A_m$ takes care of the normalization. 

The boundary conditions are derived from the sample parameters: the mean free path $\ell$, the left and right extrapolation lengths $z_{e1}, z_{e2}$ and the sample thickness $L$.

The allowed wave vectors $\kappa_m$ are obtained by solving the implicit equation
\begin{equation}
\tan \left(\kappa_m L\right) = \frac{( z_{e1} + z_{e2}) \ell \kappa_m}{z_{e1} z_{e2}\left(  \ell \kappa_m \right) ^2 - 1} \, .
\label{eq:ks}
\end{equation}

Following Ref. \cite{lagendijk1989PLA}, the phase factor $\eta_\mR$ fulfills the following equation
\begin{equation}
\tan \eta_m = z_{e1} \ell \kappa_m \, ,
\end{equation} 

and the normalization factor $A_m$ is given by 
\begin{equation}
A_m= \frac{1}{2}L - \frac{1}{2 \kappa_m} \cos ( \kappa_m L + 2\eta_\mR) \sin(\kappa_m L) \, .
\end{equation} 

The relation between eigenvalue and wavevector is $\Gamma_m = D \kappa_m^2$.

\begin{figure}[h!]
\centering
\includegraphics[scale=1]{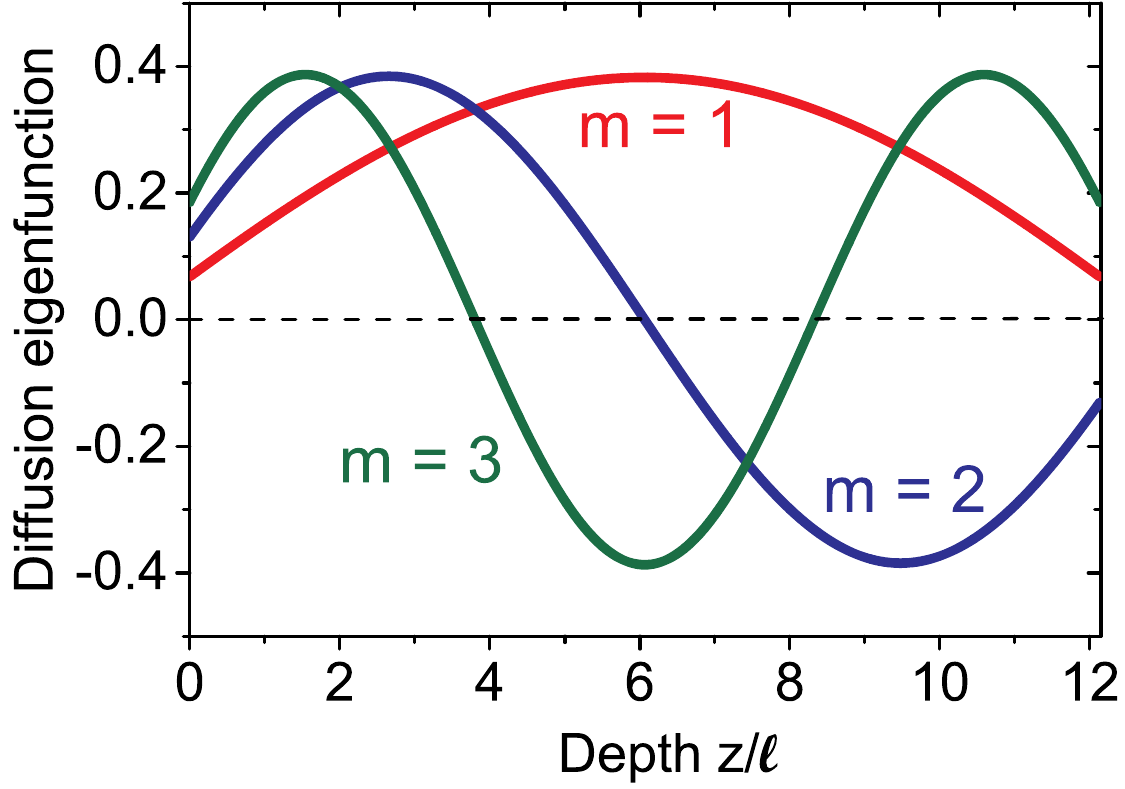}
\caption{The first three eigenfunctions of the diffusion equation plotted using equation Eq. \ref{eq:W_modes}. The wave vectors $\kappa_m$ are presented in table \ref{table:table1} and the used parameters can be found in the caption of that table. }
\label{fig:Figure1}
\end{figure}

For the exact solution of Eq. \ref{eq:ks}, see Appendix. The first ten wavevectors obtained from the solution are shown in Table \ref{table:table1} for a particular set of boundary conditions. 
In Fig. \ref{fig:Figure1}, we plot the first three eigenfunctions using Eq. \ref{eq:W_modes} with the same set of boundary conditions.

\begin{table}[t]
\centering
\begin{tabular}{ |c|c|c| } 
 \hline
  Eigenfunction index & Wavevector  \\ 
 index $m$  & $\kappa_m$   \\
 \hline \hline
 $1$ 		& 	0.229566\\ 
 $2$ 		& 	0.460711  \\ 
 $3$ 		& 	0.694591  \\ 
 $4$ 		& 	0.931747 \\ 
 $5$ 		& 	1.172199 \\ 
 $6$ 		& 	1.415643 \\
 $7$ 		& 	1.661649\\
 $8$ 		& 	1.909776 \\
 $9$ 		& 	2.159633 \\
 $10$ 		& 	2.410889 \\
 \hline
\end{tabular}
 \caption{Values of the wavevector $\kappa_m$ for the first $10$ eigenfunctions of the diffusion equation. The values were obtained by solving Eq. \ref{eq:ks}. The extrapolation lengths are $z_{e1}  = z_{e2} = z_{e} \ell = (\pi/4) \ell $ (for a 2D sample~\cite{akkermans2007book}) and L = 12.1$\ell$.}
 \label{table:table1}
\end{table}

\subsection{Reconstructing the energy density with  eigenfunctions of the diffusion equation}
The diffusion equation is of the Sturm-Liouville type so its eigenfunctions form a complete, orthogonal, set \cite{Morse1953Book}.
Therefore, any distribution of energy density within the domain of the sample boundaries can be decomposed into a sum over a finite number $M$ of eigenfunctions for a specific set of coefficients. 
Such a decomposition is only useful if the number of needed modes is small.
We express the ensemble averaged energy density $W(z)$ inside the scattering medium in terms of the normalized eigenfunctions $W_m(z)$
\begin{equation}
W(z) = \sum_{m=1}^{M} d_m W_m(z) \, , 
\label{eq:W_z}
\end{equation}
where $d_m$ are coefficients. 
We calculate the overlap integral $I_p$ of the product of $W(z)$ and $W_p(z)$ 
\begin{equation}
I_p \equiv \int_{z=0}^{z = L} dz \,  W(z) W_p(z) \, ,
\label{eq:I_def}
\end{equation} 
and substitute Eq. \ref{eq:W_z}  into Eq. \ref{eq:I_def} to obtain 
\begin{equation}
I_p = \int_{z=0}^{z = L} dz \sum_{m=1}^{M} d_m W_m(z) W_p(z)\, .
\label{eq:I_1}
\end{equation}
We use orthonormality of functions $W_p(z)$ and $W_\mR (z)$, which is expressed as 
\begin{equation}
\int_{z=0}^{z = L} dz \, W_p(z) W_m(z) \,  = \delta_{mp} \, ,
\end{equation}
where $ \delta_{\rm{mp}}$ is the Kronecker delta.
Therefore, Eq. \ref{eq:I_1} becomes
\begin{equation}
I_p = \sum_{m=1}^{M} d_m \delta_{mp}  = d_p \, .
\end{equation}

We now can write the decomposition~(\ref{eq:W_z}) as 
\begin{equation}
W(z) = \sum_{\rm{m=1}}^{M} I_{\rm{m}} W_{\rm{m}}(z) \, , 
\label{eq:W_decomp}
\end{equation}

We obtained $W(z)$ for shaped waves and different transmission eigenchannels, which we  obtained from the simulation described in Section \ref{sec:NumSamples}. The simulation results are compared to the reconstructed energy density $\Wmap$ in Section \ref{sec:results}.
In order to quantify the overlap between the reconstructed function $W_{\rm{re}}(z)$ and the numerical data $W(z)$, we use cosine distance $\rm{COSD}$~Ref[], which is defined as

\begin{equation}
\rm{COSD} \equiv 1 -  \frac{\sum\limits_{i = 1}^{N_s} W_{\rm{re}}(z_i) W(z_i)}
{ [ \sum\limits_{i = 1}^{N_s} W_{\rm{re}}^2(z_i)]^{\frac{1}{2}} 
[\sum\limits_{i = 1}^{N_s} W^2(z_i)]^{\frac{1}{2}} } \, .
\end{equation}
$N_s$ is the number of points in the numerical data. 
$\rm{COSD}$ varies between 1 and 0 and tends towards 0 as the reconstructed function fully describes the numerical data.
As a figure-of-merit of a good reconstruction, we choose the eigenfunction, which has $\rm{COSD} \approx 10^{-4}$.

\section{Numerical samples and setup}\label{sec:NumSamples}

\subsection{Numerical setup}
\begin{figure}[h!]
\centering 
\includegraphics[scale=1]{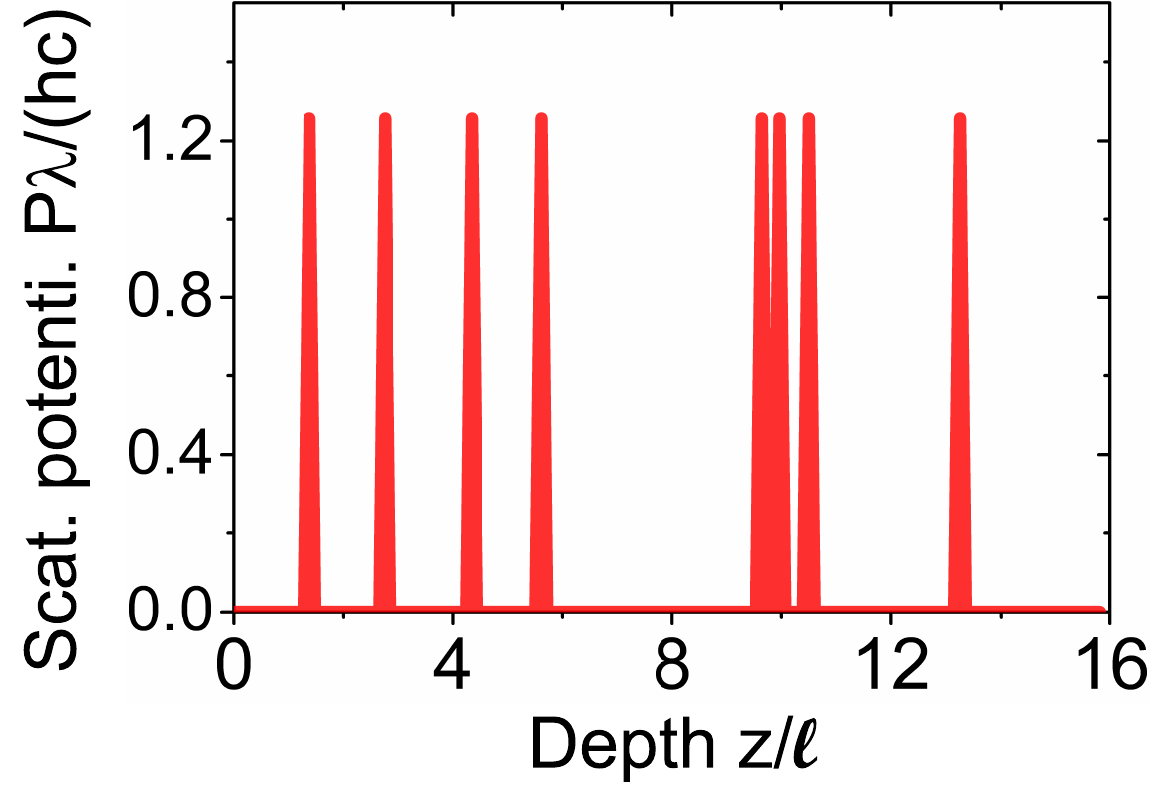}
\caption{Distribution of scattering potential $P$ of point scatters inside one configuration of the numerical samples versus the reduced sample depth $z/\ell$. The scattering potential is in the unit of $\lambda/(hc)$, $c$ is the speed of light, and h is the Planck constant.}
\label{fig:Figure2}
\end{figure}

We perform simulations of transport of monochromatic scalar waves through a waveguide containing scatterers. The waveguide is modeled as a quasi-one-dimensional (quasi-1D) system possessing $N_m=100$ transversal modes, with elastic scattering on bulk impurities. This is implemented as a 2D waveguide with 100 transversal modes, containing scatterers at random positions with a density of 0.2 scatterer per $\lambda^2$, $\lambda$ is the wavelength. 
The distribution of the scattering potential $P$ of scatterers in one waveguide configuration is shown in Fig. \ref{fig:Figure2}.
In order to calculate the energy density inside the waveguide we employ a recursive $S$-matrix formalism, where the waveguide is first divided into sub-wavelength slices, each containing only few scatterers. 
For each slice we define the $S$-matrix in a mode representation as follows \cite{Beenakker1997RMP},
\begin{equation}
S=\begin{pmatrix}
R_{L} & T^T \\ T & R_{R}
\end{pmatrix},
\end{equation}
where $R_L$ and $R_R$ are the left and right reflection matrix respectively of dimension $N_m \times N_m$, $T$ is the left-to-right transmission matrix. The right-to-left transmission matrix  in this reciprocal system is the transpose of $T$.
The $S$-matrix of each slice, of dimensions $2 N_m \times 2 N_m$, is calculated in the approximation that no recurrent scattering takes place within the thin slice. Next, the $S$-matrices of the slices are joined, including contributions of recurrent scattering, using the composition rule for $S$-matrices\cite{Ko1988PRB}. This numerically stable procedure yields the $S$-matrix of the entire waveguide and of subsections of it. From the transmission matrix we extract the singular values and vectors corresponding to all channels by singular value decomposition. The energy density inside the waveguide, for a given incident field, is calculated as follows: For a given axial coordinate $z_c$ the $S$ matrices $S_1$,$S_2$ of the waveguide sections $0<z<z_c$ and $z_c<z<L$  respectively are calculated. The field in the $z=z_c$ plane is then found by calculating recurrent scattering diagrams:
\begin{equation}
E_c=T_1 E_{in}+ R_{L2} T_1 E_{in}+ R_{R1} R_{L2} T_1 E_{in} + R_{L2} R_{R1} R_{L2} T_1 E_{in} + ...
\end{equation}
\begin{equation}
= (1+R_{L2})(1-R_{R1}R_{L2})^{-1} T_1 E_{in}.
\end{equation}
Here, $R_{L2}$ is the left reflection submatrix of $S_2$, $R_{R1}$ is the right reflection submatrix of $S_1$, and $T_1$ is the left to right transmission submatrix of $S_1$.  
Subsequently for every $z_s$ the time-averaged energy density $\varepsilon_0|E_c|^2$ is integrated over the cross section to obtain the projected energy density $W(z)$. The calculation is repeated for 8000 independently generated random waveguides from which the transmission channels are extracted. To obtain plots of the average energy density as a function of transmission, we average the energy densities of channels in a narrow band of transmission values.

\subsection{Sample characterization}

\begin{figure}[ht!]
  \centering
  \includegraphics[width=\textwidth]{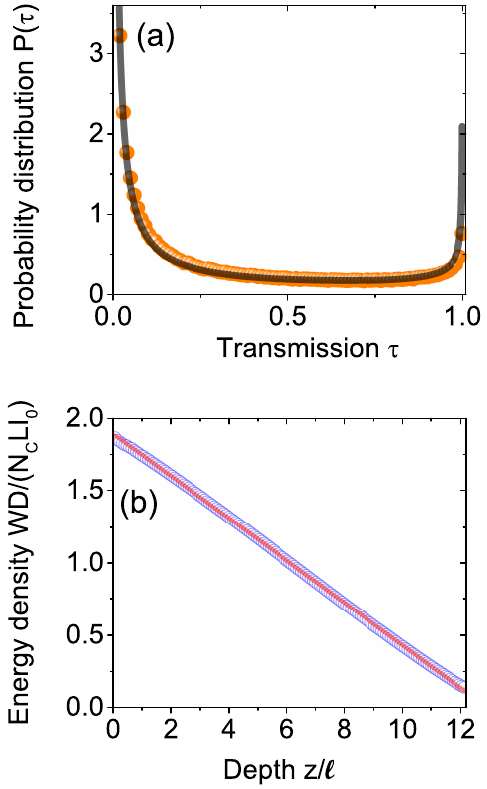}
  \caption{(a) Probability distribution of the transmission from the simulation (blue dots) and the expected Dorokhov-Mello-Pereyra-Kumar (DMPK) distribution (red curve).
(b) An equally-weighed summation over the energy density of all transmission channels (blue dots) and the expectation from diffusion theory (red line) as a function of the normalized depth $z/\ell$. 
In (a) and (b), we ensemble averaged over 8000 waveguides and the sample thickness is $L = 12.1\ell$.
 }
 \label{fig:Figure3}
\end{figure}

In order to characterize the numerical samples, we plot in Fig. \ref{fig:Figure3}a the probability distribution $P(\tau)$ as a function of the transmission $\tau$. 
The probability distribution $P(\tau)$ obtained from the simulation is bi-modal: there is a high probability for transmitting channels with transmission $\tau$ close to zero and one.
In Fig. \ref{fig:Figure3}a, we compare the Dorokhov-Mello-Pereyra-Kumar (DMPK) distribution \cite{Beenakker1997RMP} with our numerical result. 
The probability distribution of transmission for a scattering medium is expected to follow the DMPK distribution. 
The DMPK distribution agrees well with our numerical result. 
Furthermore, in order to confirm that our numerical samples are in the diffusive regime, we plot in Fig. \ref{fig:Figure3}b the equally-weighted ensemble averaged energy density versus reduced sample depth. 
The ensemble averaged energy density shows a linear decrease from the front surface towards the end surface of the sample, in agreement with the prediction of diffusion theory for a diffusive sample, see Fig. \ref{fig:Figure3}b.

\section{Results and discussions} \label{sec:results}
\subsection{Energy density of shaped waves}

\begin{figure}[ht!]
  \centering

\includegraphics[width = \textwidth]{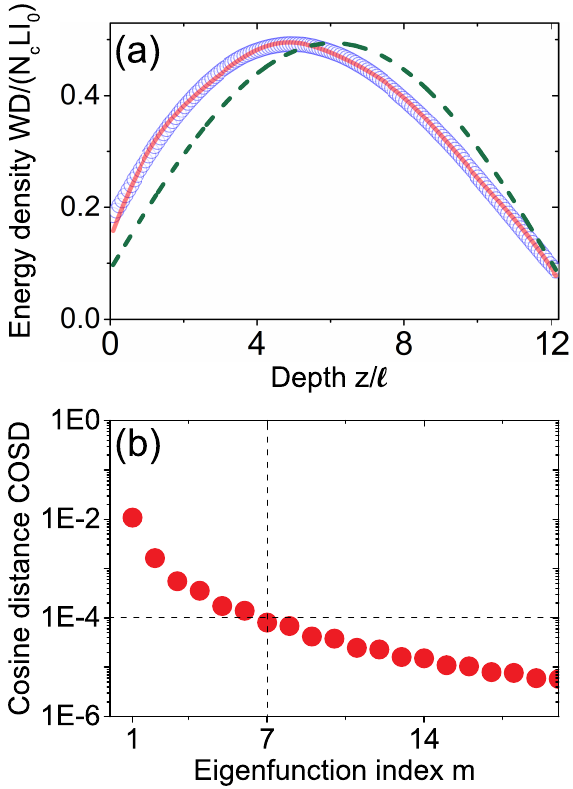} 
    \caption{  (a) Transmission-weighted averaged energy density versus normalized sample depth $z/\ell$. 
 The transmission-weighted averaged energy density is equivalent to wavefront-shaped or phase-conjugated light \cite{vellekoop2008PRL}.  
   The blue circles are the energy density obtained from simulation, the green line is the diffusion $m = 1$ eigenfunction, and the red line is the summation over the first seven eigenfunctions. 
   (b)    Cosine distance $\rm{COSD}$ versus eigenfunction index $m$.
   The vertical and horizontal dashed-lines shows position of our figure-of-merit $\rm{COSD} \approx 10^{-4}$ and the eigenfunction, which fulfills the criterion $m = 7$ respectively. 
   In both (a) and (b), the sample is the same as in Fig. \ref{fig:Figure3}.}
   \label{fig:Figure4}
\end{figure}

\begin{figure}[ht!]
\includegraphics[width = \textwidth]{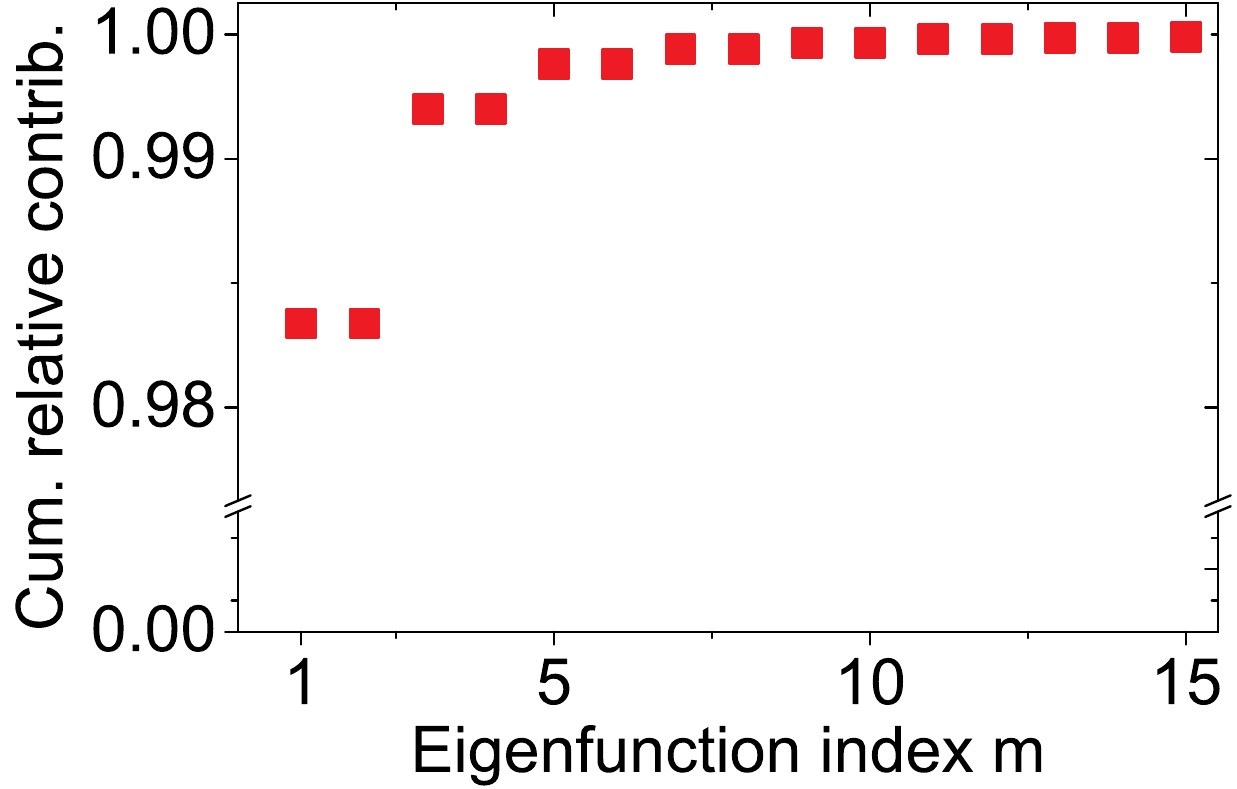} 
    \caption{  
The cumulative contribution of the eigenfunctions of the diffusion equation to the total internal energy density of shaped waves relative to a sum over the first 100 eigenfunctions versus the eigenfunction index $m$. }
   \label{fig:Figure5}
\end{figure}

We obtained the energy density of shaped waves $W_{\rm{sw}}$ as the transmission-weighted ensemble average of the energy density of transmission channels $W_n(z,\tau)$, as shown in Ref.~\cite{vellekoop2008PRL,vellekoop2015OptExp}:
\begin{equation}
W_{\rm{sw}}(z) = \sum_n^N \tau_n W_n(z,\tau) \, ,
\label{eq:W_contr}
\end{equation}
where $N$ is the number of transmission channels.
Using Eq. \ref{eq:W_contr}, we obtained the energy density of the shaped waves from the transmission-weighted averaged over the energy density of all transmission channels.
We show in Fig. \ref{fig:Figure4}a the ensemble averaged energy density distribution of shaped waves.
The energy density distribution of the shaped waves increases from the interfaces of the sample and is maximum at $z = 4.9\ell$, which is about 20$\%$ off the center of the sample ($z = 6.1\ell$).
The ensemble averaged energy density of wavefront-shaped light was first obtain in Ref. \cite{choi2011PRB} by solving the Maxwell equation using the finite-difference time-domain (FDTD) method.
Despite the different numerical calculation methods, our numerical result is closely similar to the one obtained by Choi \textit{et al} \cite{choi2011PRB}.
The exact peak position of the energy density obtain in \cite{choi2011PRB} is difficult to estimate due to the noise in the data, however, the peak position is close to the center of the sample.

We reconstruct the distribution of the energy density of shaped waves from eigenfunctions of the diffusion equation. 
From the reconstruction, we obtain the cosine distance $\rm{COSD}$ which we plot in Fig. \ref{fig:Figure4}.
In case of shaped waves, a summation over the first $M = 7$ eigenfunctions is  remarkably sufficient ($\rm{COSD} \approx 10^{-4}$) to reconstruct the energy density, see Fig. \ref{fig:Figure4}a.
Only the diffusion fundamental eigensolution $m = 1$ diffusion eigensolution has $\rm{COSD} \approx 10^{-2}$ and is not sufficient to describe the energy density profile.
The fundamental eigensolution peaks at exactly the center of the sample, in contrast to the energy density of shaped waves.

We further investigated the contribution of the eigenfunctions to the total energy density integrated along the depth of the sample relative to the contribution of the first 100 eigenfunctions. 
As a reference, we choose the first 100 eigenfunctions because the higher order eigenfunctions has a very negligible contribution.
In Fig. \ref{fig:Figure5}, we plot the contributions.
For shaped waves, the $m = 1$ contributes 98$\%$ of the total energy density inside the sample.
Remarkably, our finding about the contribution of $m = 1$ to the total internal energy density agrees very well with the heuristic model in Ref. \cite{OjambatiArxiv} that wavefront shaping \textit{predominately} couples light to the fundamental diffusion eigensolution.
The contribution of $m = 2$ is negligible compared to $m = 1$ since $m = 2$ has negative energy density in almost half of the sample depth, which cancels out the positive energy density (see Fig. \ref{fig:Figure1}).
A similar effect also happens for other even index eigenfunction. 

\subsection{Energy density of transmission channels}

\begin{figure}[h!]
\centering
\includegraphics[width = \textwidth]{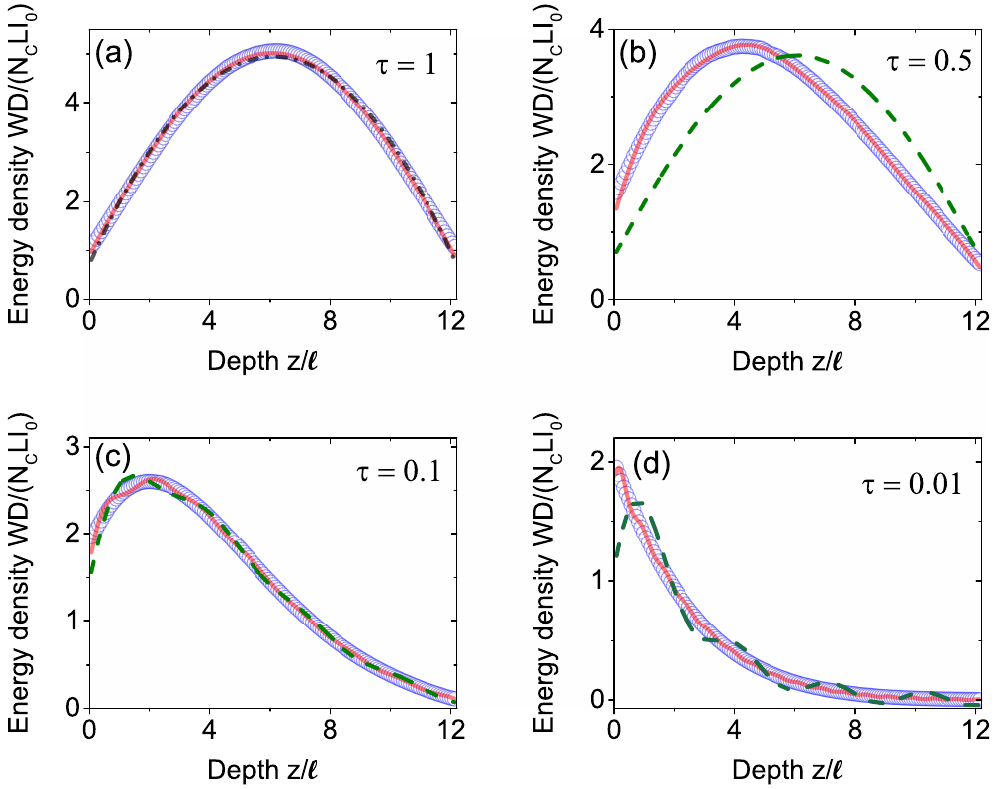}
   \caption{(a) Ensemble averaged energy density of an open channel in a scattering medium versus normalized depth $z/l_{tr}$ with transmission $\tau$ in the range 0.99 $ < \tau < 1 $. 
The blue circles are the energy density data obtained from simulation, and the red line is the fundamental diffusion eigenfunction $m = 1$. 
The black dashed-dotted curve is a parabolic fit.
(b) Ensemble averaged energy density of transmission channels in a scattering medium versus the normalized depth $z/l_{tr}$ for transmission $\tau$ in the range  $0.49 < \tau < 0.51 $. 
The blue circles are the energy density data obtained from simulation, the green line is the diffusion $m = 1$ eigenfunction, and the red line is a $M = 7$ summation of diffusion eigenfunctions. 
(c) The blue circles are obtained as in (a) and the transmission $\tau$ is in the range of $0.09 < \tau < 0.11 $. The green and the red lines are summations over $M = 8$ and $M = 16$ diffusion eigenfunctions respectively. 
(d) The blue circles are obtained as in (a) and the transmission $\tau$ is in the range: $ 0 < \tau < 0.02 $, which signifies closed channels. The green and the red lines are summations over $M = 8$ and $M = 33$ diffusion eigenfunction respectively. 
}
   \label{fig:Figure6}
\end{figure}

The ensemble averaged energy density of open transmitting eigenchannels with a transmission $\tau$ in the range $0.99  < \tau < 1 $ is shown in Fig. \ref{fig:Figure6}. 
The energy density shows a symmetric profile, which peaks in the middle of the sample. 
Our numerical result is similar to the one obtained by Davy \textit{et al} \cite{davy2015NatComm}. 
Interestingly, only the fundamental diffusion eigenfunction $m = 1$ is sufficient to reconstruct the energy density of the open channel, see Fig. \ref{fig:Figure4}a.
The fundamental diffusion eigenfunction $m = 1$ is expected to describe the open eigenchannels for the following reasons: 
Firstly, as the open eigenchannel has the highest individual transmission, the fundamental eigenfunction $(m = 1)$ as well contributes most to the total transmission as shown in Ref.\cite{OjambatiArxiv}. 
Secondly, the fundamental eigenfunction is the only physical solution with a positive energy density along the sample depth $z$, see Fig. \ref{fig:Figure1}.
Interestingly, a simplified mathematical model by Davy \textit{et al}~\cite{davy2015NatComm} described the energy density of open channels as a parabolic function. 
We note that the deviations from a parabolic shape are obvious only near the edges and the shape of the energy density curve agrees better with the cosine shape of the fundamental eigenfunction, see Fig. \ref{fig:Figure6}(a).
The $\rm{COSD}$ for the parabola is $10^{-3}$ and $10^{-5}$ for the fundamental eigenfunction.

\begin{figure}[ht]
  \centering

  \includegraphics[scale=1]{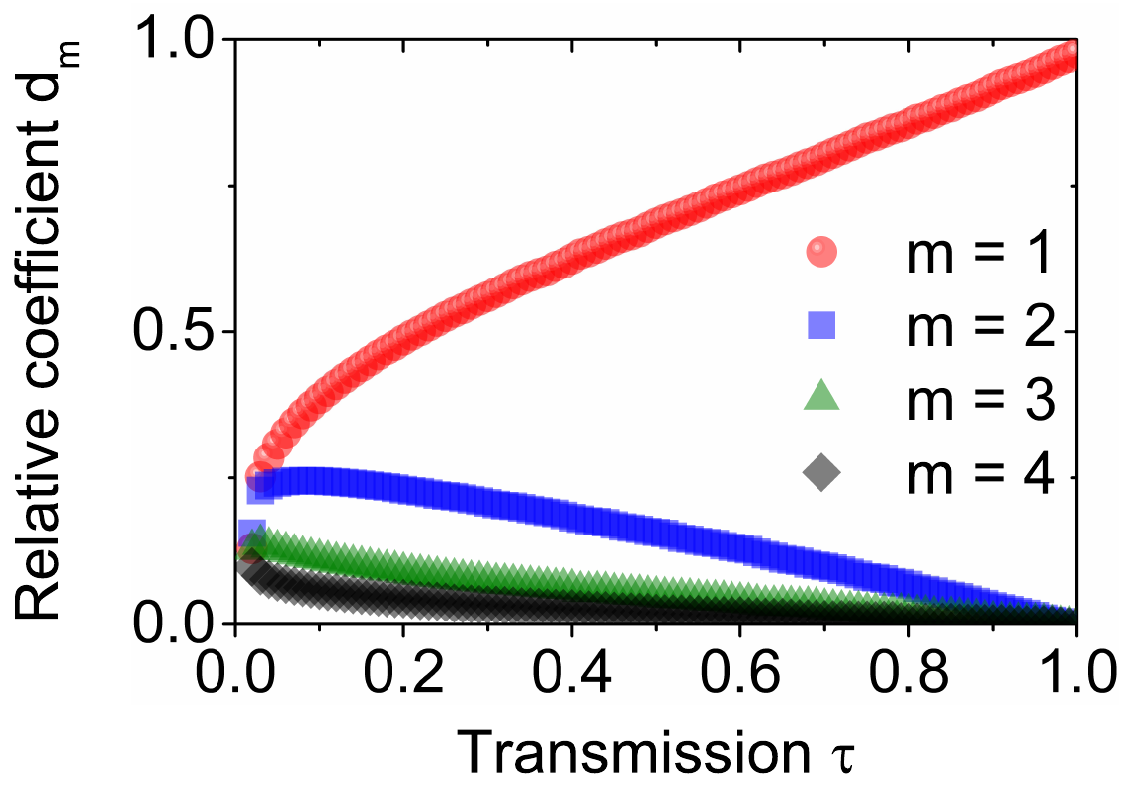}
\caption{Contribution of eigenfunctions of the diffusion equation relative to a sum of the first 100 eigenfunctions versus transmission $\tau$ for eigenfunctions $m = 1, 2, 3,$ and $4$ plotted with red circles, blue squares, green triangles, and black diamonds respectively. }
\label{fig:Figure7new}
\end{figure}

In order to describe the distribution of energy density of low-transmission channels, we also use the eigenfunctions of the diffusion equation to reconstruct the ensemble averaged energy densities of low-transmission channels obtained from simulation.
In Figs. \ref{fig:Figure6} (b - d), we show the ensemble averaged energy densities for transmission eigenchannels with transmissions $\tau = 0.5, \, 0.10$ and $0.01$.
For the eigenchannel with transmission $\tau = 0.5$, only $M = 7$ eigenfunctions of the diffusion equation are sufficient to reconstruct its energy density, $M = 16$ eigenfunctions for $\tau = 0.1$, and $M = 33$ is required for a closed channel $\tau = 0.01$. 
As the transmission $\tau$ decreases, the number of eigenfunctions sufficient to reconstruct the energy density increases. 
The increase in the number $M$ of sufficient eigenfunctions is because the asymmetry of the distribution of energy density increases as the transmission $\tau$ reduces, and likewise, the asymmetry of the eigenfunctions increases with the eigensolution index. 
Therefore, the higher index eigenfunctions contribute more to an asymmetric function.  
The contribution of the eigenfunctions with index $m$ = 1, 2, 3 and 4 relative to a summation over the first 100 eigenfunctions is shown in Fig. \ref{fig:Figure7new}.
The relative contribution of the fundamental eigenfunction of the diffusion equation $m = 1$, which is a symmetric function increases as the transmission increases, while the relative contributions of the higher index eigenfunctions, which are asymmetric functions are higher for low-transmission channels.

\subsection{Effect of sample thickness}

\begin{figure}[ht]
  \centering
  \includegraphics[scale = 2]{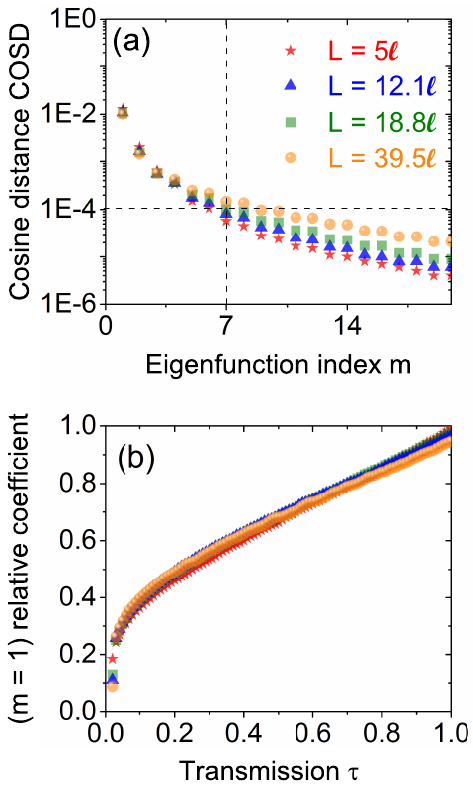}
\caption{
(a) Cosine distance $\rm{COSD}$ versus eigenfunction index $m$ for 4 samples with different thickness: $L = 5\ell, 12.1\ell, 18.8\ell$ and $39.5\ell$, which are plotted as pink stars, green squares, blue triangles, and orange circles respectively.
   The vertical and horizontal dashed-lines shows position of our figure-of-merit $\rm{COSD} \approx 10^{-4}$ and the eigenfunction, which fulfills the criterion $m = 7$ respectively.
(b) Contribution of the fundamental diffusion mode $m = 1$ relative to the sum of the first 100 eigenfunctions as a function of transmission $\tau$. }
\label{fig:Figure7}
\end{figure}

Here, we investigate the effect of the sample thickness on the number of eigenfunctions sufficient to reconstruct the energy density of shaped waves. 
We therefore perform simulation of the energy density on different samples with various sample thicknesses: $L = 5\ell, 12.1\ell, 18.8\ell$ and $39.5\ell$.
We show in Fig. \ref{fig:Figure7} the $\rm{COSD}$ versus the eigensolution index for the 4 samples. 
Interesting, all the samples have a convergence to a value of $\rm{COSD} \approx
 10^{-4}$ at the seventh eigensolution.
This convergence of the $\rm{COSD}$ shows that the summation over the first 7 eigenfunctions of the diffusion equation is sufficient to describe the energy density of shaped waves, irrespective of the sample thickness. 
There is however a slight deviation for the thickest sample $L = 39.5\ell$ and this deviation is probably due to artefacts from the simulation due to the large number of waveguide slices.

In the previous sections, we have shown that the diffusion fundamental eigenfunction of the diffusion equation $m = 1$ dominates the energy density of high-transmission channels.
We investigate if this result holds for the different samples thicknesses.
In Fig. \ref{fig:Figure7}, we show the coefficient of the $m = 1$ mode relative to the sum of the first 100 eigenfunctions versus the transmission for the 4 different thicknesses.  
The relative contribution of the $m = 1$ mode is similar for all samples, irrespective of the thickness and this is quite remarkable. 
The diffusion fundamental eigensolution dominates the energy density of high-transmission channels for all sample thicknesses.

\section{Conclusion}

We have shown that only a few eigenfunctions of the diffusion equation suffice to accurately reconstruct the distribution of the shaped energy density inside a quasi-1D scattering waveguides. 
In particular, the fundamental eigenfunction of the diffusion equation is very similar to the distribution of energy density of shaped waves.
To reconstruct the distribution of open channels only the diffusion fundamental eigenfunction $m = 1$ is sufficient.
In addition, we have shown that a few number of diffusion eigenfunctions reconstruct the energy density of low transmitting channels. 
Our results are relevant for applications that require the precise knowledge of distribution of the energy density inside scattering media. 
Such applications include as efficient light harvesting in solar cells especially in near infrared where silicon has low absorption; for enhanced energy conversion in white LEDs, which serves to reduce the quantity of expensive phosphor; for low threshold and higher output yield of random lasers; as well as in homogeneous excitation of probes in biological tissues.

\section{Acknowledgment}
We thank Pepijn Pinkse, Henri Thyrrestrup, and Hasan Yilmaz for useful discussions. This project is part
of the research program of the "Stichting voor Fundamenteel Onderzoek der Materie"
(FOM) FOM-program "Stirring of light!", which is part of the "Nederlandse Organisatie
voor Wetenschappelijk Onderzoek" (NWO). We acknowledge NWO-Vici, DARPA, and
STW.

\appendix
\section{Method of calculating the wave vectors of diffusion eigenfunctions}
In Eq. \ref{eq:ks}, we show an implicit equation, which defines the allowed wave vectors of the diffusion eigenfunctions. 
The equation is again stated here:
\begin{equation}
\tan \left(\kappa_m L\right) = \frac{(z_{e1} + z_{e2}) \ell \kappa_m}{z_{e1}z_{e2}\left(  \ell \kappa_m \right) ^2 - 1} \, .
\label{eq:ks_append}
\end{equation}
We define 
\begin{equation}
f(\kappa_m) = f_1(\kappa_m) - f_2(\kappa_m) \, ,
\end{equation}
where 
\begin{equation}
f_1(\kappa_m)  \equiv \tan \left(\kappa_m L\right)
\label{eq:f1}
\end{equation}
and 
\begin{equation}
f_2(\kappa_m)  \equiv  \frac{(z_{e1} + z_{e2}) \ell \kappa_m}{z_{e1}z_{e2}\left( \ell \kappa_m \right) ^2 - 1} \, .
\label{eq:f2}
\end{equation}
 Our goal is to find the wave vectors $\kappa_m^{\rm{r}}$ that fulfills the condition
 \begin{equation}
f(\kappa_m^{\rm{r}}) = 0 \, . 
\label{eq:rooteq}
\end{equation}

\begin{figure}[ht]
  \centering

  \includegraphics[scale=1]{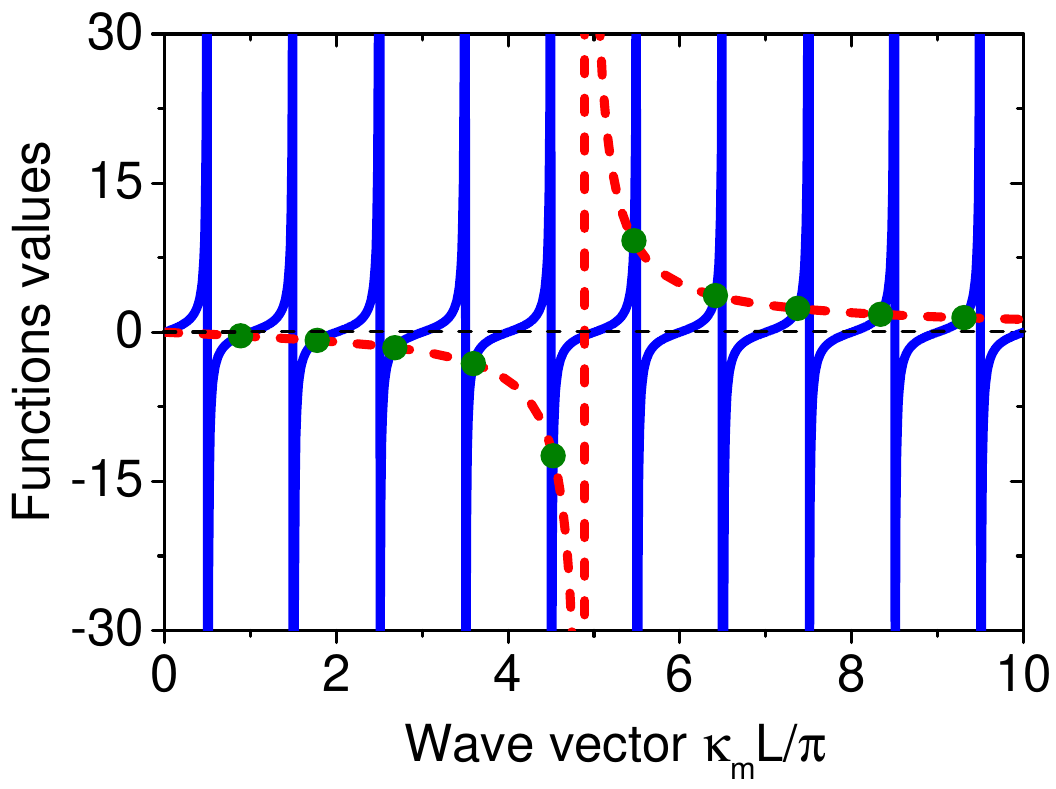}
\caption{The values of functions $f_1$ (blue line) and $f_2$ (red dashed line) in Eqs. \ref{eq:f1} and \ref{eq:f2}. as a function of reduced diffusion wave vector $\kappa_m L / \pi$. The green circles are the roots of Eq. \ref{eq:ks_append}. }
\label{fig:FigureA1}
\end{figure}

In Fig. \ref{fig:FigureA1}, we plot the two functions $f_1$ and $f_2$. 
The condition in Eq. \ref{eq:rooteq} is fulfilled by the wave vectors $\kappa_m^{\rm{r}}$ at the intersection of the two functions $f_1$ and $f_2$, as indicated in Fig. \ref{fig:FigureA1}.
We found the wave vectors $\kappa_m^{\rm{r}}$ by dividing function $f$ into domains with size $\pi/L$. 
In most of the domains, only one root is expected to be in each of the domains, expect the domain where function $f_2$ has a divergence, which has two roots.
In each domain, we use a Simple Bisection method to search the domain for wave vectors that fulfill the desired condition. 
Each domain is divided into two equal parts. 
The function $f$ is first evaluated for left part of the domain and then check if there is any root present. 
If there is no root present, the left domain is further divided into smaller subdomains and each subdomains is searched for the root. 
We divide the domains until the size of the subdomain is $10^{-12}$. 
If there is no root in all the subdomains of the left domain, then, the right domain is searched.
This way, all the desired roots are found, iteratively.

\end{document}